# Pressure-induced topological phase transition in noncentrosymmetric elemental Tellurium

T. Ideue[a, 1], M. Hirayama[b], H. Taiko[a], T. Takahashi[c], M. Murase[c],

T. Miyake[d], S. Murakami[e,f], T. Sasagawa[c], and Y. Iwasa[a,b]

[a] Quantum-Phase Electronics Center (QPEC) and Department of Applied Physics,
The University of Tokyo, Tokyo 113-8656, Japan

[b] RIKEN Center for Emergent Matter Science (CEMS), Wako, Saitama 351-0198, Japan

[c] Laboratory for Materials and Structures, Tokyo Institute of Technology, Yokohama, Kanagawa 226-8503, Japan

[d] Research Center for Computational Design of Advanced Functional Materials, National Institute of Advanced Industrial Science and Technology, Tsukuba 305-8568, Japan.

[e]Department of Physics, Tokyo Institute of Technology, 2-12-1 Ookayama, Meguro-ku, Tokyo 152-8551, Japan

[f]Tokodai Institute for Element Strategy, Tokyo Institute of Technology, 2-12-1 Ookayama, Meguro-ku, Tokyo 152-8551, Japan.

**Abstract**

Recent progress in understanding the electronic band topology and emergent topological properties encourage us to reconsider the band structure of well-known materials including elemental substances. Controlling such a band topology by external field is of particular interest from both fundamental and technological view point. Here we report the pressure-induced topological phase transition from a semiconductor to a Weyl semimetal in elemental tellurium probed by transport measurements. Pressure variation of the periods of Shubnikov-de Haas oscillations, as well as oscillations phases, shows an anomaly around the pressure theoretically predicted for topological phase transition. This behavior can be well understood by the pressure-induced band deformation and resultant band crossing effect. Moreover, effective cyclotron mass is reduced toward the critical pressure, potentially reflecting the emergence of massless linear dispersion. The present result paves the way for studying the electronic band topology in well-known compounds and topological phase transition by the external field.

**Significance**

**Topological Weyl semimetal, which is a gapless semimetallic phase protected by symmetry, generally appears by band gap closing in noncentrosymmetric semiconductors. So far, there has been only a limited number of reports of such a topological phase transition so that many basic aspects have been remained unexplored. Here we report the topological phase transition under pressure in elemental tellurium with noncentrosymmetric chiral crystal structure. Our result represents that elemental tellurium is one of the model systems for the topological phase transition from a semiconductor to a Weyl semimetal, opening the avenue for studying the topological phase transition by the external field and resultant topological properties.**

Topological phase of matter is a central issue in recent condensed matter physics. It is characterized by the nontrivial electronic band topology which causes the emergent electrical, magnetic, and optical properties of solids (1, 2). One of the important topics in this field is the search for the phase control between topologically trivial phase and topologically nontrivial phase. Such a topological phase transition is generally realized by changing the chemical composition of compounds (3-7) or by applying the external field such as the magnetic field, electric field (8) or pressure (9-11). Although it can be directly probed by the angle resolved photoemission spectroscopy in the former case, optical or transport measurements and comparison with the theoretical calculations are necessary for verifying the topological phase transition induced by external fields.

An elemental substance is an ideal platform to study characteristic properties of topological phases and topological phase transition since it has a simple crystal or band structure and we can easily compare it with theoretical calculations (12-17). Since the spin-orbit interaction often plays the key role for the band inversion and the emergence of the topologically nontrivial phase, elemental substance with large spin-orbit interaction can be a potential candidate. For example, topological nature of electronic bands has been discussed in Bi, Sb and As so far (12-15). Tellurium is also a potential elemental substance with strong spin-orbit interaction. Figure 1A shows the crystal structure of tellurium. Helical chains along the $z$ axis are arranged via weak van-der-Waals interaction in a hexagonal array. Depending on the chirality of helical chains, it belongs to the space group $P3_121$ or $P3_221$. At the ambient pressure, Tellurium is a $p$-type semiconductor in which electronic and optical properties are mainly dominated by the valence and conduction band around H point in the hexagonal Brillouin zone (Fig. 1B). Previous studies clarified the characteristic camel-back-like valence band dispersion with minimum energy $E_c$ at H point (Fig. 1C left) and that strong spin-orbit interaction should be taken into account to explain the transport and optical properties (18).

Observation of the current induced magnetization due to the chiral structure has been reported (19) and Te is known also as a good thermoelectric material with high figure of merit due to its valley degeneracy and band nesting (20).

Recently, topological nature of Te band is attracting renewed interest due to its noncentrosymmetric crystal structure and strong spin-orbit interaction (21-23). First principle calculation predicts that elemental Tellurium shows the topological phase transition under pressure (23). Under the application of pressure, characteristic camel-back-like valence band dispersion first disappears (Fig. 1C middle), turning into the normal parabolic band dispersion. With increasing the pressure, band gap continuously shrinks so that valence band and conduction band finally touch at the points around the H point (Fig. 1C right), leading to the topological transition from a semiconductor to a Weyl semimetal. (In Supporting Information, we show the detailed band evolution under pressure.) In Weyl semimetal phase, band crossing points (Weyl points) are stable against the perturbation and will not disappear to open the gap (i.e., topologically protected). This argument of topological transition under pressure is consistent with the general discussion for noncentrosymmetric crystals based on the symmetry (24).

In this study, we have probed the topological phase transition in elemental Tellurium through electric transport, especially focusing on the Shubnikov-de Haas (SdH) oscillations. Periods of SdH oscillations, as well as phases, sensitively change by pressure, reflecting the pressure-induced band deformation. In a sample with low carrier density, periods of SdH oscillations show an anomalous jump around $P = 0.5$ GPa, which is consistent with the theoretically-predicted pressure-induced Lifshitz transition (See Supporting Information). Furthermore, non-monotonous pressure dependences of SdH oscillation periods and phases have been observed in all samples irrespective of carrier density, which can be well understood by the characteristic Fermi surface deformation and Berry phase change during the topological phase transition. Our results represents that elemental tellurium is one of the model systems which show the topological phase transition from a semiconductor to a Weyl semimetal and SdH oscillations can be a sophisticated probe to detect the band deformation and resultant topological phase transition.

**Results and Discussion**

In Figs. 2A and 2B, we show the magnetoresistance ($R_{zz}$) and Hall effect ($R_{xz}$) of sample 1 under various pressures. In this work, we mainly applied the magnetic field perpendicular to the natural growth surface (*zx*-plane) and current direction is fixed to be the *z*-axis. (We also studied pressure dependence of SdH oscillations in another configuration ($B \parallel I \parallel z$) in the Supporting Information.) Carrier density $n_{\text{Hall}}$ of sample 1 estimated from Hall effect is $n_{\text{Hall}} = 2.7 \times 10^{16}$ cm$^{-3}$. (Carrier number studied in this work range from $n_{\text{Hall}} \sim 10^{16}$ cm$^{-3}$ to $10^{17}$ cm$^{-3}$.) Clear SdH oscillations have been observed at *T* = 2K in both $R_{zz}$ and $R_{xz}$, which are apparently dependent on the pressure. Figure 2C represents the oscillating component of $R_{zz}$ calculated by subtracting the polynomial function. Maximum and minimum positions are linearly fitted in the index plots (Fig. 2D). In this index plot, we assigned minimum of $R_{zz}$ as an integer. Cross-sectional areas of the Fermi surface ($S_F$) are calculated from the periods of the oscillating components (or slopes of index plots) by the following relation and are plotted as a function of the pressure in Fig. 2E.

$$S_F = \frac{2\pi e}{\hbar \Delta(1/B)},$$

where $e$, $\hbar$, and $\Delta(1/B)$ are electron charge, Plank's constant and the period of the oscillation, respectively.

At the ambient pressure, $S_F$ calculated from the oscillation period is $0.526 \times 10^{-3}$ Å$^{-2}$, indicating that Fermi energy of sample 1 locates above the minimum of the camel-back-like valence band dispersion ($E_c$) (See Supporting Information.). As we increase the pressure, each peak in oscillations first shows the systematic change (Fig. 2C region 1) but suddenly splits into two peaks around $P = 0.5$ GPa (Fig. 2C region 2). At this pressure, $S_F$ estimated from the period of SdH oscillations is almost doubled as shown in Fig. 2E (region 1 and 2). This split of SdH oscillation and anomalous jump of $S_F$ is considered to be the signature of pressure-induced Lifshitz transition. As predicted by calculations (Fig. 1C), characteristic camel-back-like valence band dispersion disappears under pressure. Fermi level of sample 1, first locating above the minimum of the camel-back-like valence band dispersion ($E_c$), passes through the $E_c$ during the systematic pressure-induced band deformation, which cause the Fermi surface topology change and resultant Lifshitz transition. In Fig. 2F, we show the pressure variations of the calculated 3D Fermi surface in which Lifshitz transition is illustrated in the first step.

After this Lifshitz transition, SdH oscillations again show the systematic change with pressure, showing another anomaly around $P = 2$ GPa (Fig. 2E); when the pressure is below 2 GPa, $S_F$ decrease with pressure (region 2) while it turns upward above $P = 2$ GPa (region 3). This anomaly is considered to be caused by the band touching predicted by calculations. Before the band touching occurs, Fermi surface is becoming more 3D-like and isotropic by applying the pressure so that Fermi surface is elongated along the $k_x$ and $k_y$ direction while it shrinks along the $k_z$ direction. Therefore, cross-sectional area of the Fermi surface perpendicular to the magnetic field ($k_z$-$k_x$ plane) decreases mainly reflecting the elongation along the $k_y$ direction (second and third schematics in Fig. 2F). On the other hand, once the conduction band and valence band touches, the shape of the Fermi surface becomes anisotropic again since the original conduction band minimum keeps decreasing under pressure and causes the deformation of the Fermi surface as it gets close to the Fermi energy (second and third schematics in Fig. 1C and third and fourth schematics in Fig. 2F). Thus, the anomaly of $S_F$ around $P = 2$ GPa represents the band touching and resultant topological phase transition from a semiconductor to a Weyl semimetal.

We note that Hall coefficient is also affected by the pressure in sample 1 (Fig. 2B), which indicates the carrier increase under pressure. It might be because that carriers trapped by the defect levels are released under pressure, causing the increase of the carrier density toward $P = 2$Gpa (Fig. 2B). However, the increase of the carrier density cannot explain the observed decrease of $S_F$ so that pressure dependence of $S_F$ does not come from the carrier density change but can be mainly attributed to the pressure-induced band deformation. It is also noted that Hall coefficient is almost unchanged above 2 GPa. In Supporting Information, we discuss the phase of SdH oscillations (or intercept of the index plot), which show the crossover behavior from a semiconductor to a Weyl semimetal, offering the important insight into the change of the band dispersion or Berry phase originating from spin texture (25-31)

We also studied the samples with different Fermi energies at the ambient pressure. Figures 3A-F show $R_{zz}$ (A), $R_{xz}$ (B), oscillating components of $R_{zz}$ (C), index plot (D) under various pressures, and pressure dependence of $S_F$ (E), and calculated 3D Fermi surface (F) for sample 2. Carrier density $n_{\text{Hall}}$ of sample 2 estimated from Hall effect is $n_{\text{Hall}} = 4.1 \times 10^{17}$ cm$^{-3}$. According to the oscillation period, the Fermi level of sample 2 is estimated to locate below $E_c$ at the ambient pressure (See Supporting Information.). Differently from sample 1, $S_F$ does not show an anomalous jump around $P = 0.5$ GPa (Fig. 3E). This means there is no pressure-induced Lifshitz transition in sample 2, being consistent with $E_F < E_c$. Nevertheless, pressure dependence of $S_F$ still shows the upturn around $P = 2$ GPa (Fig. 3E) similarly to the sample 1, indicating that topological phase transition from a semiconductor to a Weyl semimetal (Fig. 3F) is successfully probed also in sample 2. We have confirmed that similar behavior can be observed in other two samples with Fermi energy below $E_c$ at the ambient pressure (Sample 3 and sample 4. See Supporting Information.). Intercept $\delta$ in the index plot shows crossover behavior around $P = 2$ GPa also in sample 2, corroborating the above scenario of topological phase transition (See Supporting Information).

Finally, we discuss the pressure variation of the effective mass of sample 2. In Figure 4A, oscillating components of sample 2 under $P = 3$ GPa at various temperatures are displayed. Magnitude of the oscillations decreases with increasing the temperature, which is plotted in Fig. 4B. According to the following Lifshitz-Kosevich formula for a three-dimensional system (32), we can estimate the effective mass $m^*$ from the temperature dependence of the oscillation amplitude $\Delta R_{zz}/R_{zz}$,

$$\Delta R_{zz}/R_{zz} \propto (\hbar\omega_C/2E_F)^{\frac{1}{2}} \exp(-2\pi^2 k_B T_D/\hbar\omega_C) \frac{2\pi^2 k_B T/\hbar\omega_C}{\sinh(2\pi^2 k_B T/\hbar\omega_C)}.$$

Here, $k_B$, $T_D$, and $\omega_C$ are Boltzmann's constant, Dingle temperature and cyclotron frequency $\omega_C = \frac{eB}{m^*}$, respectively. We calculate the effective mass at each pressure from the temperature dependence of $\Delta R_{zz}/R_{zz}$ (Fig. 4B) and plot values normalized by free electron mass $m_0$ as a function of the pressure in Fig. 4C. Effective cyclotron mass is reduced by applying the pressure, reflecting the transition from massive band dispersion at the ambient pressure to the massless linear band dispersion in Weyl semimetal phase. Pressure variation of $m^*$ is settled around $P = 2$ GPa similarly to $\delta$, which almost coincides with the anomaly point of $S_F$, possibly because that band deformation is suppressed after the topological phase transition. From the fitting of temperature variation of SdH oscillations by Lifshitz-Kosevich formula, we also obtained the Fermi velocity $v_F = 4.0 \times 10^4$ m/s and quantum lifetime $\tau_Q = 1.4 \times 10^{-13}$ s at $P = 3$ GPa. Transport lifetime defined by $\tau_{\text{transport}} = \mu\hbar k_{\text{F}}/ev_{\text{F}}$ is estimated as $\tau_{\text{transport}} = 8.5 \times 10^{-13}$ s, which is order of magnitude larger than $\tau_{\text{Q}}$. It is noted that $\tau_{\text{transport}}$ generally larger than $\tau_{\text{Q}}$, since $\tau_{\text{transport}}$ measures backscattering process that relaz the current while $\tau_{\text{Q}}$ is sensitive to all processes that cause Landau level broadening. (30).

In conclusion, we have successfully probed the pressure-induced topological phase transition from a semiconductor to a Weyl semimetal in elemental tellurium. Present results indicate that electric transport including SdH oscillations is a powerful probe to detect the topological phase transition, especially under external field such as the pressure. In view of wealthy families of materials whose topological nature is not fully unveiled, our findings will open the way to rediscover the topologically nontrivial band and topological phase transition in well-known materials.

## Materials and Methods

### Sample preparations.

Single crystals of Te were prepared by a physical vapor transport (PVT) technique. Elemental Te (6N) grains of ~0.5 g were sealed in an evacuated quartz tube (~ φ11 mm x φ9 mm x 120 mm), and the ampule was put in a two-heating-zone tube furnace (33). Temperatures at the source (Te grains) and the growth zone were set to 450 $^{\circ}$C and 360 $^{\circ}$C, respectively. After ~ 70 h, single crystals of Te with ~3 mm length having a hexagonal prism shape as similar to a quartz crystal were obtained.

### Transport measurements.

Resistivity and Hall effect have been measured with the use of a physical property measurement system (Quantum Design, Inc.). Hydrostatic pressure is applied by using a piston-cylinder pressure cell and Daphne 7474 (pressure-transmitting medium). We applied the magnetic field perpendicular to the natural growth surface (*zx*-plane) and current direction is parallel to the *z*-axis.

### First principle calculation.

Our calculations are based on the density function theory (DFT) in the local density approximation (LDA). We calculate the fully-relativistic electronic structure by a first-principles code QMAS (Quantum MAterials Simulator) based on the projector augmented-wave method. The plane-wave energy cutoff is set to 40 Ry, and the 6×6×6 *k*-mesh is employed. We evaluate the self-energy correction in the GW approximation (GWA) (34, 35) using the full-potential linear muffin-tin orbital code (36, 37). In the GWA, we neglect the spin-orbit interaction (SOI). The 6×6×6 *k*-mesh is sampled and 51×2 unoccupied conduction bands are included, where ×2 is the spin degrees of freedom. We construct 9×2 maximally localized Wannier functions (MLWF's) originating from the *5p* orbitals and diagonalize the fully-relativistic Hamiltonian with the GW self-energy correction expressed in the MLWF basis. Experimental structures for the trigonal phase are used in the present study. At pressures between available experimental data, we determine the Hamiltonian in the MLWF basis by linear interpolation.

**ACKNOWLEGEMENTS**

This research was supported by the following grants; JSPS Grant-in-Aid for Specially Promoted Research (No. JP25000003), Grant-in-Aid for Challenging Research (Exploratory) (No. JP17K18748), Grant-in-Aid for Scientific Research on Innovative Areas "Topological Materials Science" (No. 18H04216), Grants-in-Aid for Scientific Research (B) (No. 16H03847), and JST CREST project (No. JPMJCR16F2).

Author contributions: T. I. and Y. I. conceived and designed the experiments. T. T., M. M., T. S. performed the growth and characterization of the bulk samples. T. I. and H. T. measured transport properties. M. H., T. M., and S. M. performed the theoretical calculations. T. I, M. H., T. S. and Y. I. wrote the manuscript. All authors led the physical discussions and contributed to improve the manuscript.

[1]To whom correspondence should be addressed. Email: ideue@ap.t.u-tokyo.ac.jp

This article contains suppoting information.

**References**

1. Qi XL, and Zhang SC (2011) Topological insulators and superconductors. *Rev Mod Phys* 83: 1057-1109.

2. Armitage NP, Mele EJ, Vishwanath A (2018) Weyl and Dirac semimetals in three-dimensional solids. *Rev Mod Phys* 90: 015001.

3. Hsieh D, et al. (2008) A topological Dirac insulator in a quantum spin Hall phase. *Nature* 452: 970-974.

4. Hsieh D, et al. (2009) A tunable topological insulator in the spin helical Dirac transport regime. *Nature* 460: 1101-1105.

5. Dziawa P, et al. (2012) Topological crystalline insulator states in $Pb_{1-x}Sn_xSe$. *Nat Mater* 11: 1023-1027.

6. Xu SY, et al. (2014) Topological Phase Transition and Texture Inversion in a Tunable Topological Insulator. *Science* 332: 560-564.

7. Brahlek M, et al. (2012) Topological-Metal to band-Insulator Transition in $(Bi_{1-x}In_x)_2Se_3$ Thin Films. *Phys Rev Lett* 109: 186403.

8. Qian X, Liu J, Fu L, Li J (2014) Quantum spin Hall effect in two-dimensional transition metal dichalcogenides. *Science* 346: 1344-1347.

9. Xi X, et al. (2013) Signatures of Pressure-Induced Topological Quantum Phase Transition in BiTeI. *Phys Rev Lett* 111: 155701.

10. Ideue T, et al. (2014) Pressure variation of Rashba spin splitting toward topological transition in the polar semiconductor BiTeI. *Phys. Rev. B* 90: 161107(R).

11. Liang T, et al. (2017) A pressure-induced topological phase with large Berry curvature in Pb$_{1-x}$Sn$_x$Te. *Sci Adv* 3: e1602510.

12. Drozdov IK, et al. (2014) One-dimensional topological edge states of bismuth bilayers. *Nat Phys* 10: 664-669.

13. Schindler F, et al. (2018) Higher-order topology in bismuth. *Nat Phys* 14: 918-924.

14. Yao G, et al. (2013) Evolution of Topological Surface States in Antimony Ultra-Thin Films. *Sci Rep* 3: 2010.

15. Zhang P, et al. (2017) Topologically Entangled Rashba-Split Shockley States on the Surface of Grey Arsenic. *Phys Rev Lett* 118: 046802.

16. Castro Neto AH, Guinea F, Peres NMR., Novoselov KS, Geim AK (2009) The electronic properties of graphene. *Rev Mod Phys* 81: 109-162.

17. Kim J, et al. (2015) Observation of tunable band gap and anisotropic Dirac semimetal state in black phosphorus. *Science* 349: 723-726.

18. Gerlach E, Grosse P (1979) The Physics of Selenium and Tellurium. *Springer Series in Solid-State Sciences* 13 (Springer-Verlag Berlin Heidelberg New York).

19. Furukawa T, Shimokawa Y, Kobayashi K, Ito T (2017) Observation of current-induced bulk magnetization in elemental tellurium. *Nat Commun* 8: 954.

20. Lin S, et al. (2016) Tellurium as a high-performance elemental thermoelectric. *Nat Commun* 7: 10287.

21. Agapito LA, Kioussis N, Goddard IIIWA, Ong NP (2013) Novel Family of Chiral-Based Topological Insulators: Elemental Tellurium under Strain. *Phys Rev Lett* 110: 176401.

22. Nakayama K, et al. (2017) Band splitting and Weyl nodes in trigonal tellurium studied by angle-resolved photoemission spectroscopy and density functional theory. *Phys Rev B* 95: 125204.

23. Hirayama M, Okugawa R, Ishibashi S, Murakami S, Miyake T (2015) Weyl Node and Spin Texture in Trigonal Tellurium and Selenium. *Phys Rev Lett* 114: 206401.

24. Murakami S, Hirayama M, Okugawa R, Miyake T (2017) Emergence of topological semimetals in gap closing in semiconductors without inversion symmetry. *Sci Adv* 3: e1602680.

25. Luk'yanchuk IA, Kopelevich Y (2004) Phase Analysis of Quantum Oscillations in Graphite. *Phys Rev Lett* 93: 166402.

26. Qu DX, Hor YS, Xiong J, Cava RJ, Ong NP (2010) Quantum Oscillations and Hall Anomaly of Surface States in the Topological Insulator $Bi_2Te_3$. *Science* 329: 821-824.

27. Murakawa H, et al. (2013) Detection of Bery's Phase in a Bulk Rashba Semiconductor. *Science* 342: 1490-1493.

28. Doiron-Leyraud N, Szkopek T, Pereg-Barnea T, Proust C, Gervais G (2015) Berry phase in cuprate superconductors. *Phys Rev B* 91: 245136.

29. Ideue T, et al. (2015) Thermoelectric probe for Fermi surface topology in the three-dimensional Rashba semiconductor BiTeI. *Phys Rev B* 92: 115144.

30. Liang T, et al. (2015) Ultrahigh mobility and giant magnetoresistance in the Dirac semimetal $Cd_3As_2$, *Nat Mater* 14: 280-284.

31. Xiong, J, et al. (2015) Evidence for the chiral anomaly in the Dirac semimetal $Ba_3Bi$. *Science* 350: 413-416.

32. Shoenberg D (1984) "Magnetic Oscillations in Metals" *Cambridge University Press, Cambridge, England*.

33. Iwasa A, et al. (2010) Environmentally Friendly Refining of Diamond-Molecules via the Growth of large Single Crystals. *Crystal Growth and Design* 10: 870-873.

34. Hedin L (1965) New Method for Calculating the One-Particle Green's Function with Application to the Electron-Gas Problem. *Phys Rev* 139: A796.

35. Hedin L, Lundqvist S (1969) *Solid State Physics* Vol. 23, eds. Ehrenreich, H., Seitz, F., and Turnbull, D. (Academic, New York).

36. Van Schilfgaarde M, Kotani T, Faleev SV (2006) Adequacy of Approximations in GW Theory. *Phys Rev B* 74: 245125.

37. Miyake T, Aryasetiawan F (2008) Screened Coulomb Interaction in the Maximally Localized Wannier Basis. *Phys Rev B* 77: 085122.

**Figure Legends**

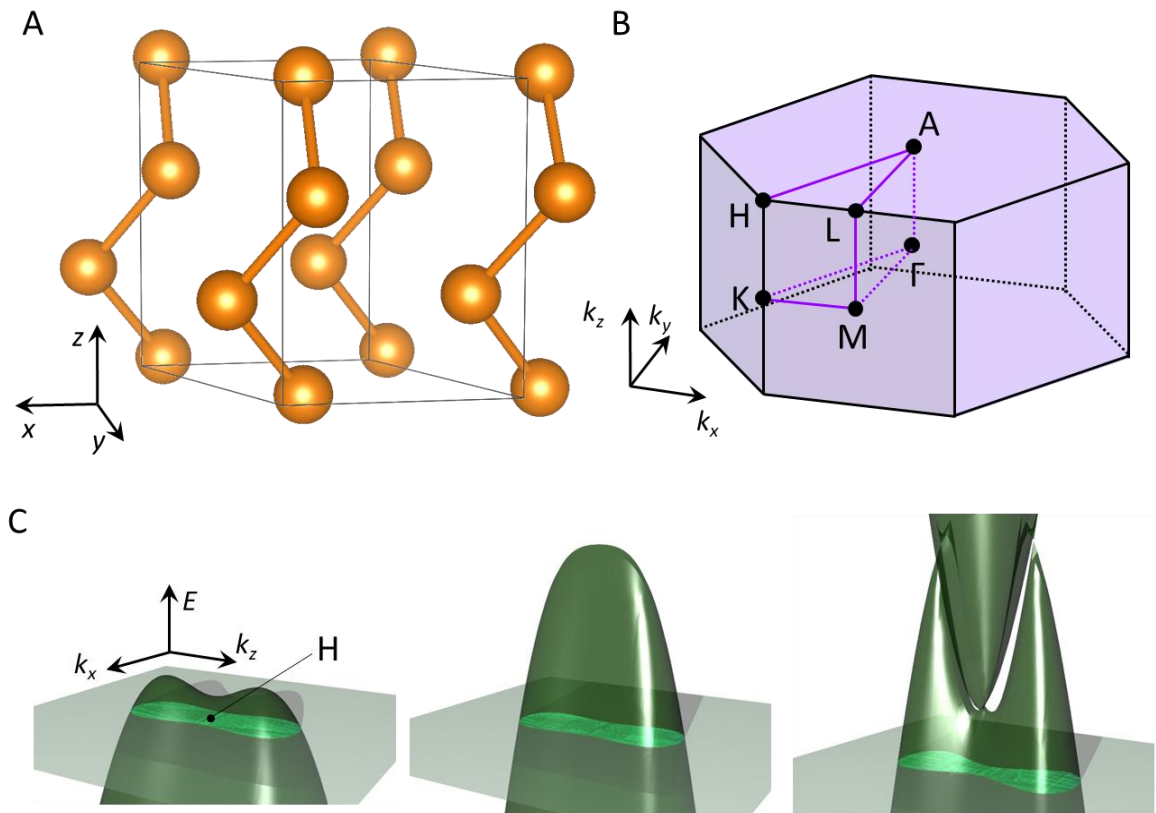

**Fig. 1.** Pressure-induced band deformation and topological phase transition of elemental tellurium. (*A*) Crystal structure of Te. Helical chains along the *z* axis are weakly coupled by van der Waals force. (*B*) Brillouin zone of Te. Electronic band around H points are dominant in electronic and optical properties of Te. (*C*) First principle calculations of Te band under pressure ($P = 0$ GPa (left), $P = 1.88$ GPa (middle), and $P = 2.62$ GPa (right), respectively) and cross-section of the Fermi surface (shaded region) when the magnetic field is applied perpendicular to the cleavage plane (*zx*-plane). Topological phase transition from a semiconductor to a Weyl semimetal is predicted around $P = 2$ GPa.

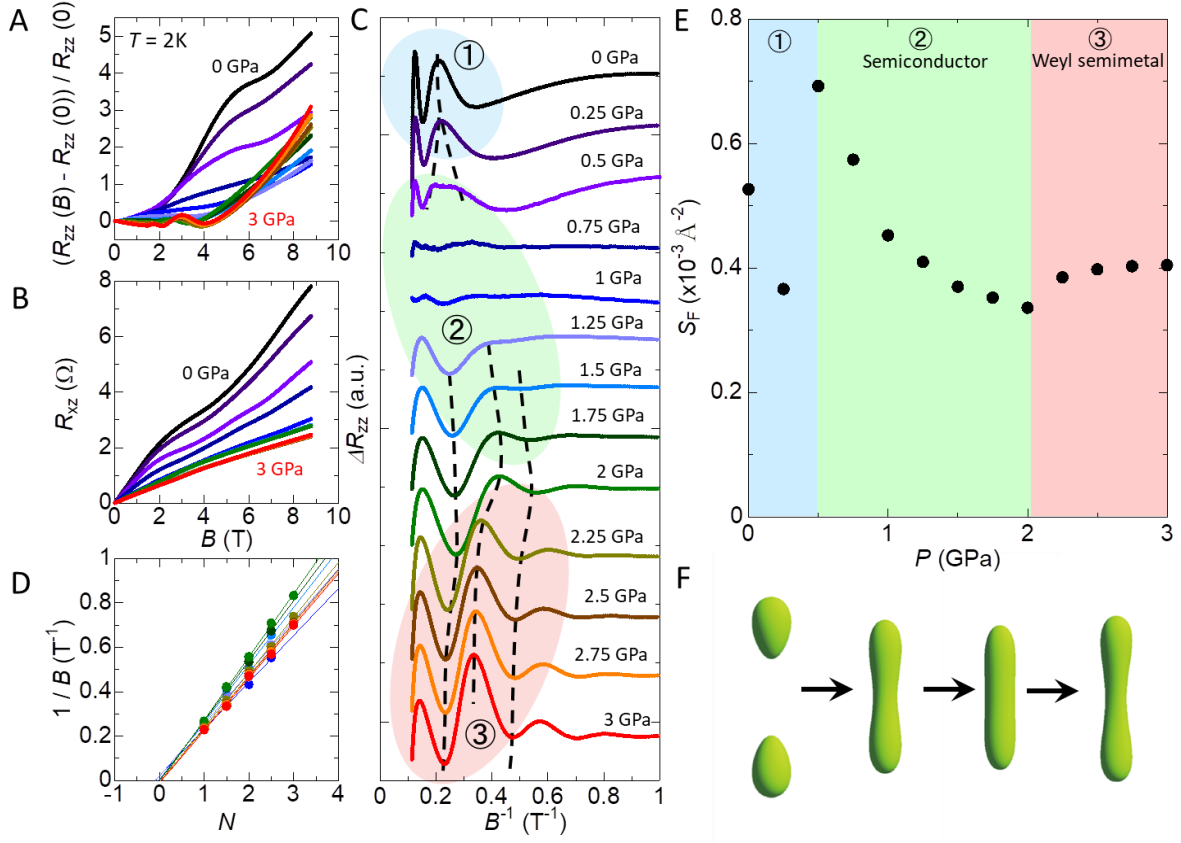

**Fig. 2.** SdH oscillations of sample 1 ($E_F > E_c$). (*A* and *B*) Magnetoresistance (A) and Hall effect (*B*) of sample 1 under various pressures. Clear SdH oscillations have been observed at $T$ = 2K. (*C*) Oscillating components of the resistance calculated by subtracting the polynomial background from A. (*D*) Landau index plot of sample 1. We chose the minimum of the resistance as an integer index. (*E*) Pressure dependence of the cross-sectional area of Fermi surface estimated from the periods of oscillations. $S_F$ shows the anomalous jump around $P$ = 0.5 GPa and upturn at $P$ = 2 GPa. (*F*) Pressure evolution of the calculated there-dimensional Fermi surface for the sample with $E_F$ = 1 meV at ambient pressure. From the left figure, $P$ = 0 GPa, $P$ = 0.6 GPa, $P$ = 1.22 GPa, and $P$ = 2.16 GPa, respectively.

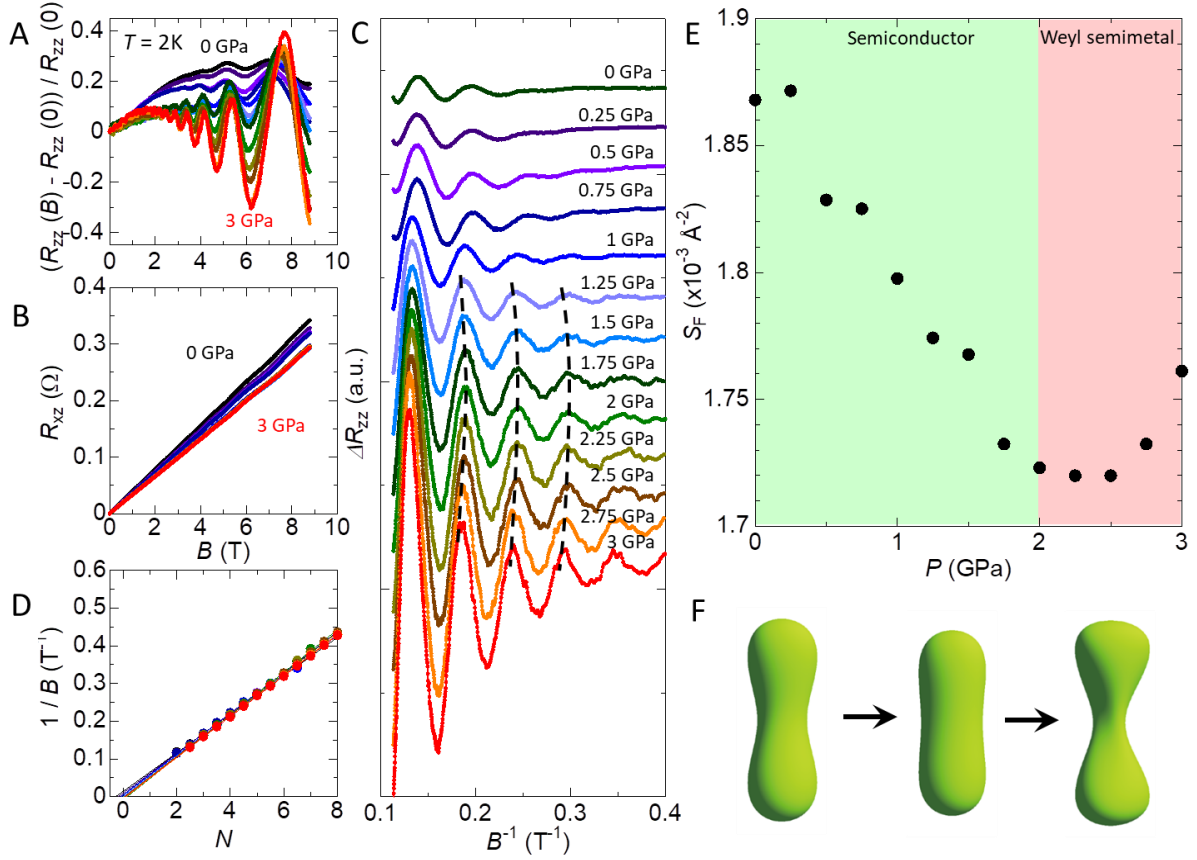

**Fig. 3.** SdH oscillations of sample 2 ($E_F < E_c$). (*A* and *B*) Magnetoresistance (A) and Hall effect (*B*) of sample 2 at $T = 2K$ under various pressure. (*C*) Oscillating components of the resistance. (*D*) Landau index plot of sample 2. (*E*) Pressure dependence of the cross-sectional area of Fermi surface estimated from the periods of oscillations. $S_F$ shows the anomaly around $P = 2$ GPa similarly to sample 1 (Fig. 2E). (*F*) Pressure evolution of the calculated there-dimensional Fermi surface for the sample with $E_F = 5$ meV at ambient pressure. From the left figure, $P = 0$ GPa, $P = 1.22$ GPa, and $P = 2.62$ GPa, respectively.

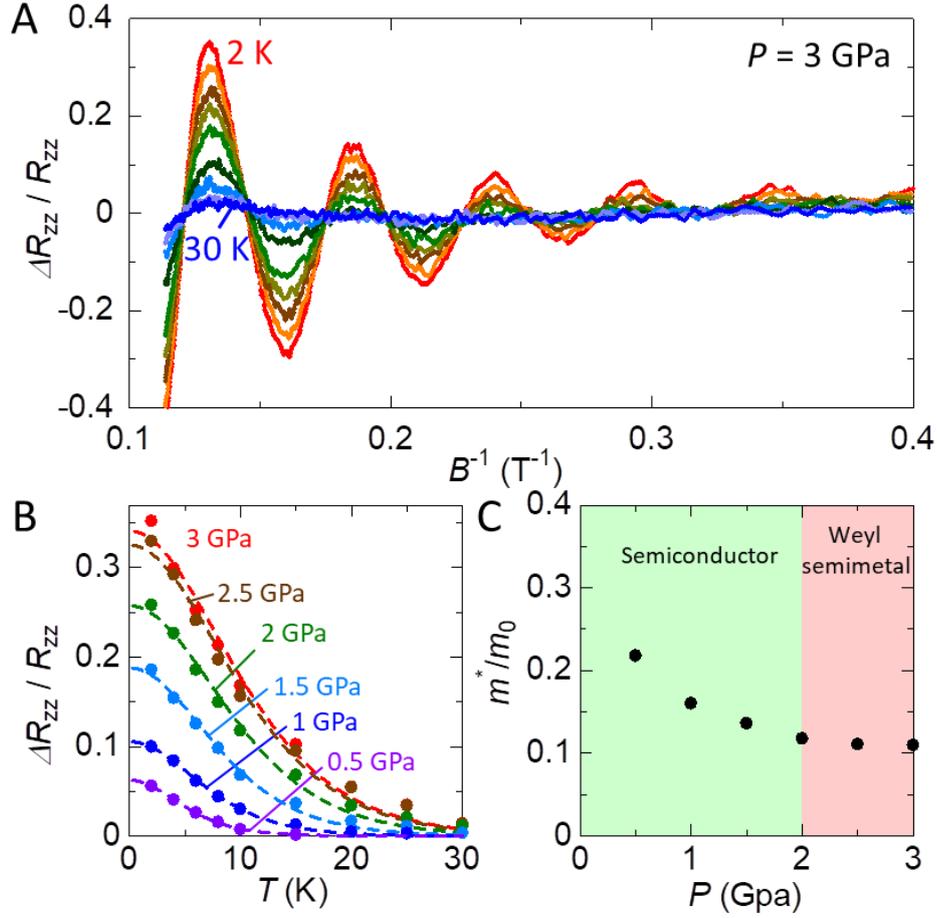

**Fig. 4.** Pressure variation of the effective mass. (*A*) Oscillating component of sample 2 under $P = 3$ GPa at various temperatures. (*B*) Temperature dependence of amplitude of SdH oscillations under various pressures. Dashed line represents the fitting by the Lifshitz-Kosevich formula. (*C*) Pressure variation of the effective mass $m^*$ normalized by free electron mass $m_0$. $m^*$ decreases with pressure, showing the convergent behavior toward $P = 2$ Gpa.

# Supporting Information

**Pressure-induced topological phase transition in non-centrosymmetric elemental Tellurium**

T. Ideue[a], M. Hirayama[b], H. Taiko[a], T. Takahashi[c], M. Murase[c],

T. Miyake[d], S. Murakami[e,f], T. Sasagawa[c], and Y. Iwasa[a,b]

1. **Fermi level position for each sample**
2. **Sample information**
3. **Detailed behavior of the pressure-induced band evolution**
4. **Pressure variation of $\delta$ in sample 1 and 2**
5. **$\rho_{zz}$ oscillations v.s. $\sigma_{zz}$ oscillations in sample 1 and 2**
6. **SdH oscillations in sample 3 and 4**
7. **SdH oscillations in sample 5 and 6 ($B \parallel I \parallel z$)**

## 1. Fermi level position for each sample

We discuss the Fermi level position of each sample by comparing the calculated cross-sectional area of Fermi surface at ambient pressure with those obtained in the experiment. Figure S1 shows the band structure at ambient pressure obtained by the first principle calculation. Energy level of the minimum of the camel-back-like valence band dispersion ($E_c$) at H point is around $E$ = -2 meV measured from the valence band top. Shape of the calculated there-dimensional Fermi surface at $E_F$ = -2 meV actually shows the critical behavior (nearly connected two Fermi surfaces, Fig. S1B). Calculated cross-sectional area of the Fermi surface at $E_F$ = 2 meV ($S_F$ = 0.592× $10^{-3}$ Å$^{-2}$) is larger than that of sample 1, while double of it is smaller than those of sample 2, 3, and 4. Thus, we can conclude that Fermi level of sample 1 locates above $E_c$ while those of sample 2, 3, and 4 are below $E_c$.

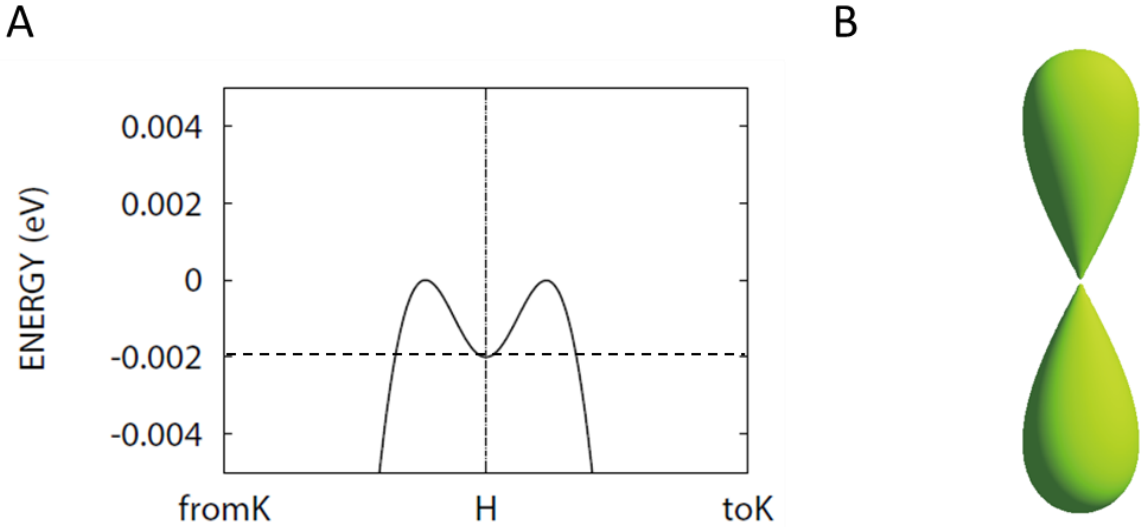

**Fig. S1.** Fermi surface around $E_c$ = -2 meV. (*A*) First principle calculation of Te band at ambient pressure. Minimum of the camel-back-like valence band dispersion ($E_c$) locates around $E_c$ = -2 meV. (*B*) Calculated there-dimensional Fermi surface for $E_F$ = -2 meV.

## 2. Sample information

In Table S1, we show the carrier density $n_{\text{Hall}}$ estimated from Hall effect, mobility $\mu$ ($P$ = 0 GPa) and $\mu$ ($P$ = 3 GPa) at $T$ = 2 K, cross-sectional area of the Fermi surface $S_F$ calculated from period of SdH oscillations, and corresponding Fermi wave number $k_F$ or effective carrier density $n_{\text{eff}}$ under ambient pressure for each sample. In the calculation of $k_F$ or $n_{\text{eff}}$, we assumed the simple isotropic linear band dispersion and several relations between parameters (based on the discussion in Nat. Mater. **14** 280 (2015)). We note that such an assumption is no longer valid in this system with large anisotropy and rather complex band structure. Nevertheless, $n_{\text{Hall}}$ and $n_{\text{eff}}$ show consistent values. From the temperature variation of SdH oscillations in sample 2 (Fig.4), we obtained the Fermi velocity $v_F = 4.0 \times 10^4$ m/s and quantum lifetime $\tau_Q = 1.4 \times 10^{-13}$ s of ample 2 at $P$ = 3 GPa. Transport lifetime defined by $\tau_{\text{transport}} = \mu \hbar k_F / e v_F$ is estimated as $\tau_{\text{transport}} = 8.5 \times 10^{-12}$ s, which is order of magnitude larger than $\tau_Q$. ($\tau_{\text{transport}}$ generally larger than $\tau_Q$, since $\tau_{\text{transport}}$ measures backscattering process that relaz the current while $\tau_Q$ is sensitive to all processes that cause Landau level broadening.

We also show the calculated carrier density $n_{\text{cal}}$ and $S_F$ at $E_F = -1, -3$, and $-5$ meV in Table S2. Experimentally obtained values of $n$ and $S_F$ in Table S1 show the similar values at $E_F = -1$ meV for sample 1 and $E_F = -5$ meV for sample 2~4, respectively (Fig. S2). Thus, we discuss the qualitative behavior of the pressure-induced topological phase transition by using the calculated values for $E_F = -1$ meV (sample 1, Fig. 2F) and $E_F = -5$ meV (sample 2, Fig. 3F).

| Sample | $n_{\text{Hall}}$ (cm$^{-3}$) | $\mu$ (cm$^2$/Vs) ($P$ = 0 GPa) | $\mu$ (cm$^2$/Vs) ($P$ = 3 GPa) | $S_F$ (10$^{-3}$ Å$^{-2}$) | $k_F$ (Å$^{-1}$) | $n_{\text{eff}}$ (cm$^{-3}$) |
|---|---|---|---|---|---|---|
| 1 | 2.7 × 10$^{16}$ | 6400 | 33000 | 0.5 | 0.0126 | 6.7 × 10$^{16}$ |
| 2 | 4.13 × 10$^{17}$ | 3100 | 31000 | 1.87 | 0.0244 | 4.9 × 10$^{17}$ |
| 3 | 2.0 × 10$^{17}$ | 1000 | — | — | — | — |
| 4 | 6.25 × 10$^{17}$ | 1000 | — | 1.76 | 0.0237 | 4.5 × 10$^{17}$ |

**Table S1.** Carrier density $n_{\text{Hall}}$ estimated from Hall effect, mobility $\mu$ ($P$ = 0 GPa) and $\mu$ ($P$ = 3 GPa) at $T$ = 2 K, cross-sectional area of the Fermi surface $S_F$ calculated from period of SdH oscillations, and corresponding Fermi wave number $k_F$ or effective carrier density $n_{\text{eff}}$ under ambient pressure for each sample.

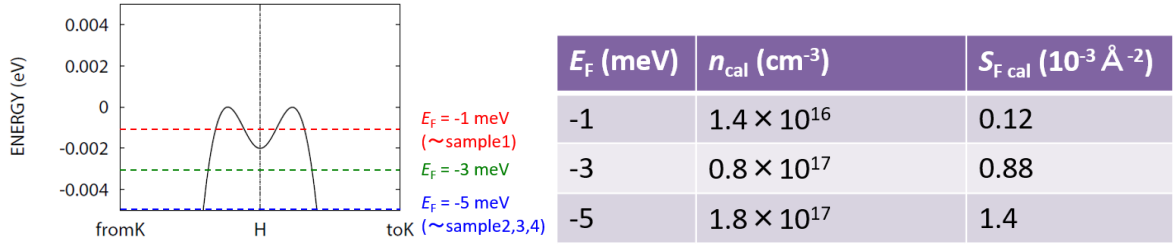

| $E_F$ (meV) | $n_{\text{cal}}$ (cm$^{-3}$) | $S_{F\,\text{cal}}$ (10$^{-3}$ Å$^{-2}$) |
|---|---|---|
| -1 | 1.4 × 10$^{16}$ | 0.12 |
| -3 | 0.8 × 10$^{17}$ | 0.88 |
| -5 | 1.8 × 10$^{17}$ | 1.4 |

**Fig. S2. (left)** Schematics of Fermi level ($E_F = -1, -3$, and $-5$ meV.).

**Table S2. (right)** Calculated carrier density $n_{\text{cal}}$ and $S_F$ at $E_F = -1, -3$, and $-5$ meV.

## 3. Detailed behavior of the pressure-induced band evolution

In Figs. S3 A-L, we show the detailed pressure-induced band evolution.

In Figs. A, C, E, G, I, K we show both conduction and valence bands. Figs. B, D, F, H, J, L are the magnified view of valence band.) We calculate all band structures by the fully-relativistic Hamiltonian with the GW self-energy correction.

Conduction band continuously decreases by applying the pressure and finally touches around $P = 2$ GPa. As shown in Figs. S3 B and D, Fermi level of sample 1 ($E_F \sim -1$meV) will cross the minimum of the camel-back-like valence band dispersion between $P = 0$ GPa and $P = 0.6$ GPa, leading to Lifshitz transition in this sample.

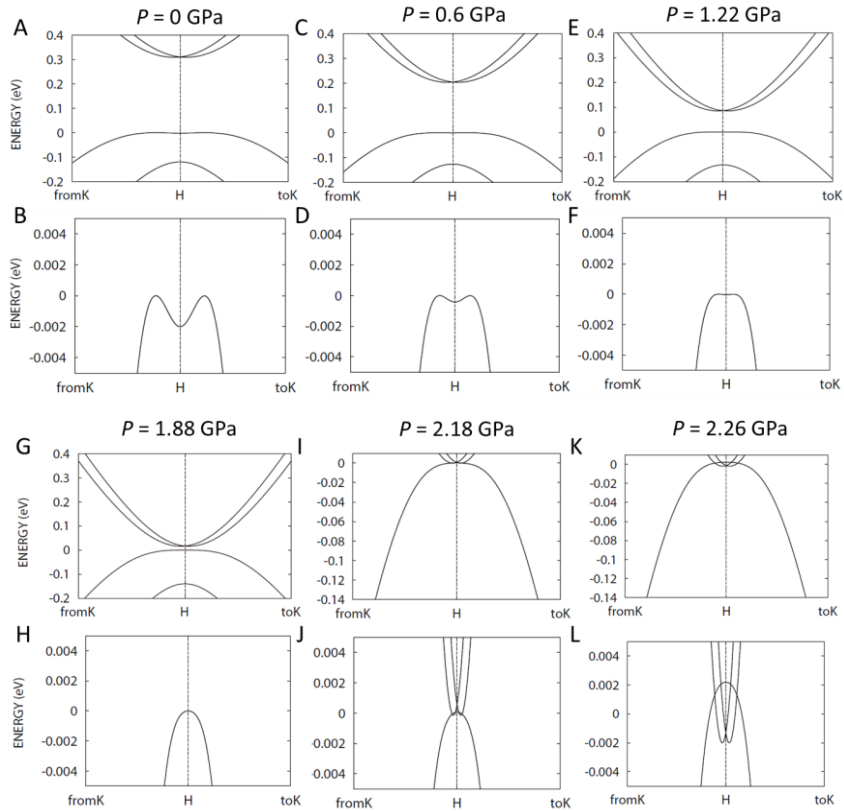

**Fig. S3.** Pressure-induced band evolution of Te. $P = 0$ GPa (A and B), 0.6 GPa (C and D), 1.22 GPa (E and F), 1.88 GPa (G and H), 2.18 GPa (I and J), and 2.26 GPa (K and L).

## 4. Pressure variation of $\delta$ in sample 1 and 2

In addition to the period of oscillations, which provides us with information of the cross-sectional area of Fermi surface, the phase of SdH oscillations (or intercept of the index plot) also offers the important insight into the band dispersion or Berry phase originating from spin texture. Figures S4 A (C) and B (D) show the magnified view of the index plot and pressure dependence of $\delta$ of sample 1 (sample 2). $\delta$ shows crossover behavior from -0.2 in the semiconducting phase to 0.1 in the Weyl semimetal phase, possibly reflecting the change of band dispersion. However, the value of $\delta$ are not 0 (parabolic dispersion) nor 1/2 (Dirac dispersion), varying from -0.2 to 0.1, potentially due to the complex spin texture/dispersion of valence band in Te. It is also noted that analysis of the phase of SdH oscillations $\delta$ in topological semimetal is not simple as discussed in Nat. Mater. **14** 280 (2015) or Science **350** 413 (2015). Quantitative argument of $\delta$ and its relation to the complex spin texture should be further studied in the future.

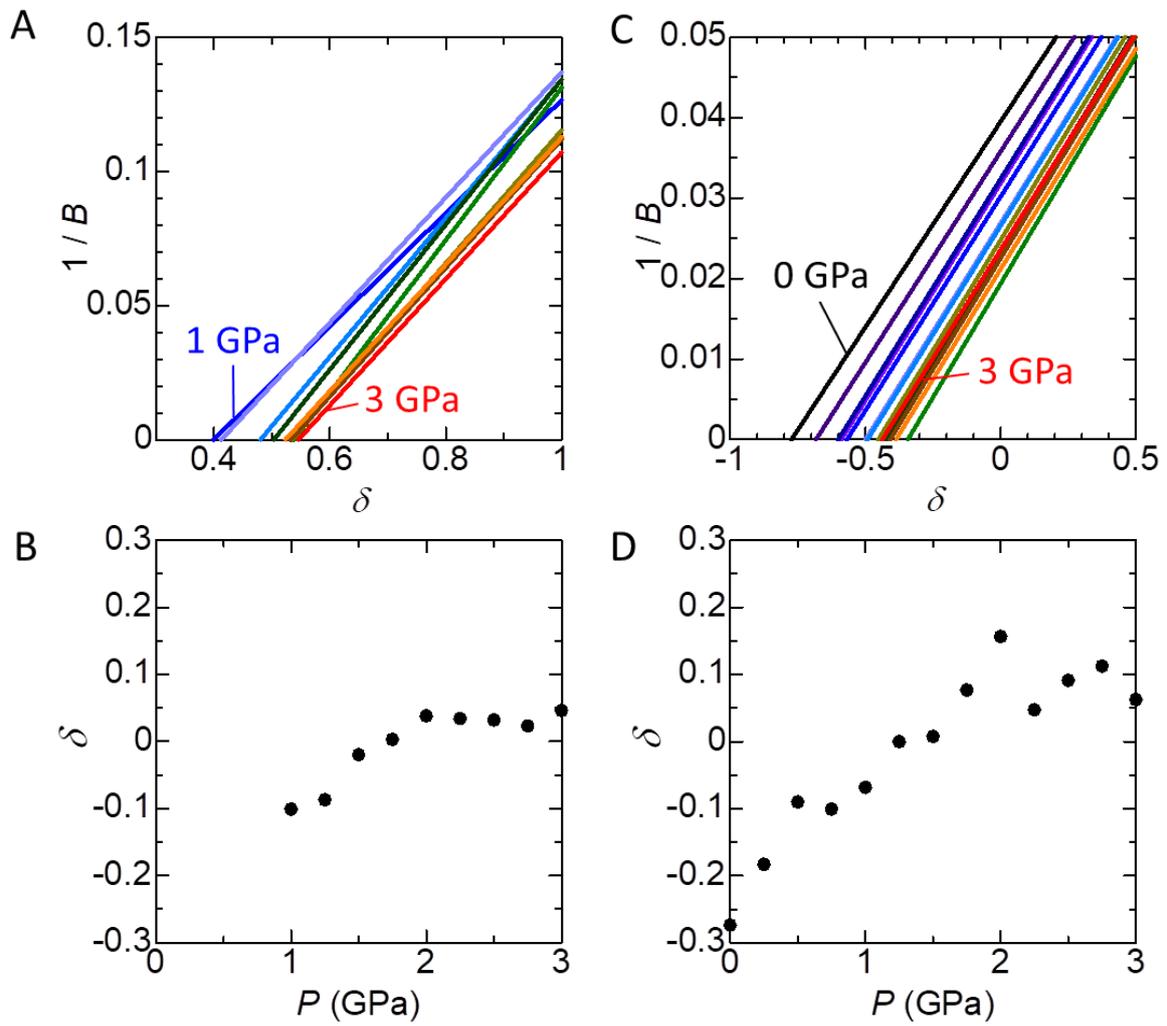

**Fig. S4.** Pressure variation of $\delta$ in sample 1 and 2. (*A*, *B*) Magnified view of the index plot (A) and pressure dependence of $\delta$ (B) of sample 1. (*C*, *D*) Magnified view of the index plot (C) and pressure dependence of $\delta$ (D) of sample 2.

## 5. $\rho_{zz}$ oscillations v.s. $\sigma_{zz}$ oscillations in sample 1 and 2

In the main text, we argue the $\rho_{zz}$ oscillations of sample 1 and sample 2. In this section, we compare $\sigma_{zz}$ oscillations with those in $\rho_{zz}$. In Figs. S5 and Figs. S6, we show the oscillating components of $\sigma_{zz}$ (A), index plots (B), and pressure variation of the horizontal axis intercept $\delta$ in index plots (C) of sample 1 (Figs. S5) and sample 2 (Figs. S6), respectively. Since Hall angle is large enough in both samples, oscillating components in $\sigma_{zz}$ show similar behavior as those in $\rho_{zz}$; i.e., $\Delta\sigma_{zz} \sim \Delta\rho_{zz}$. Following the index plots of $\Delta\rho_{zz}$ in the main text, we assigned minimum of $\Delta\sigma_{zz}$ as an integer in Fig. S5B and Fig. S6B $\delta$ shows crossover behavior from semiconducting phase to Weyl semimetal phase, which is almost consistent with that of $\rho_{zz}$ oscillations.

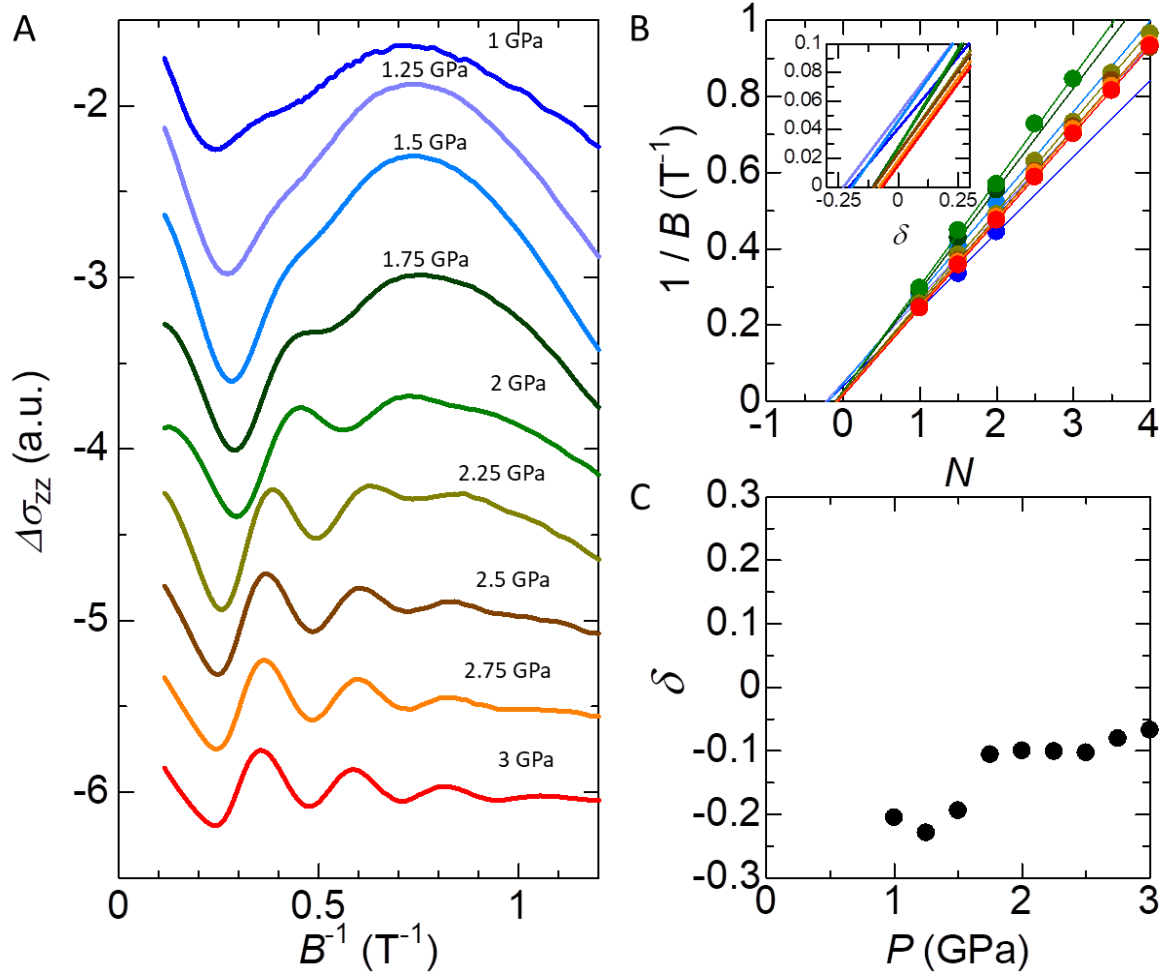

**Fig. S5.** $\sigma_{zz}$ oscillations of sample 1. (*A*) SdH oscillations of $\sigma_{zz}$ in sample 1. (*B*) Index plot for $\sigma_{zz}$ oscillations in sample 1. (*C*) Pressure variation of the horizontal axis intercept $\delta$ in the index plot. $\delta$ shows the similar crossover behavior toward $P = 2$ GPa as that for $\rho_{zz}$ oscillations.

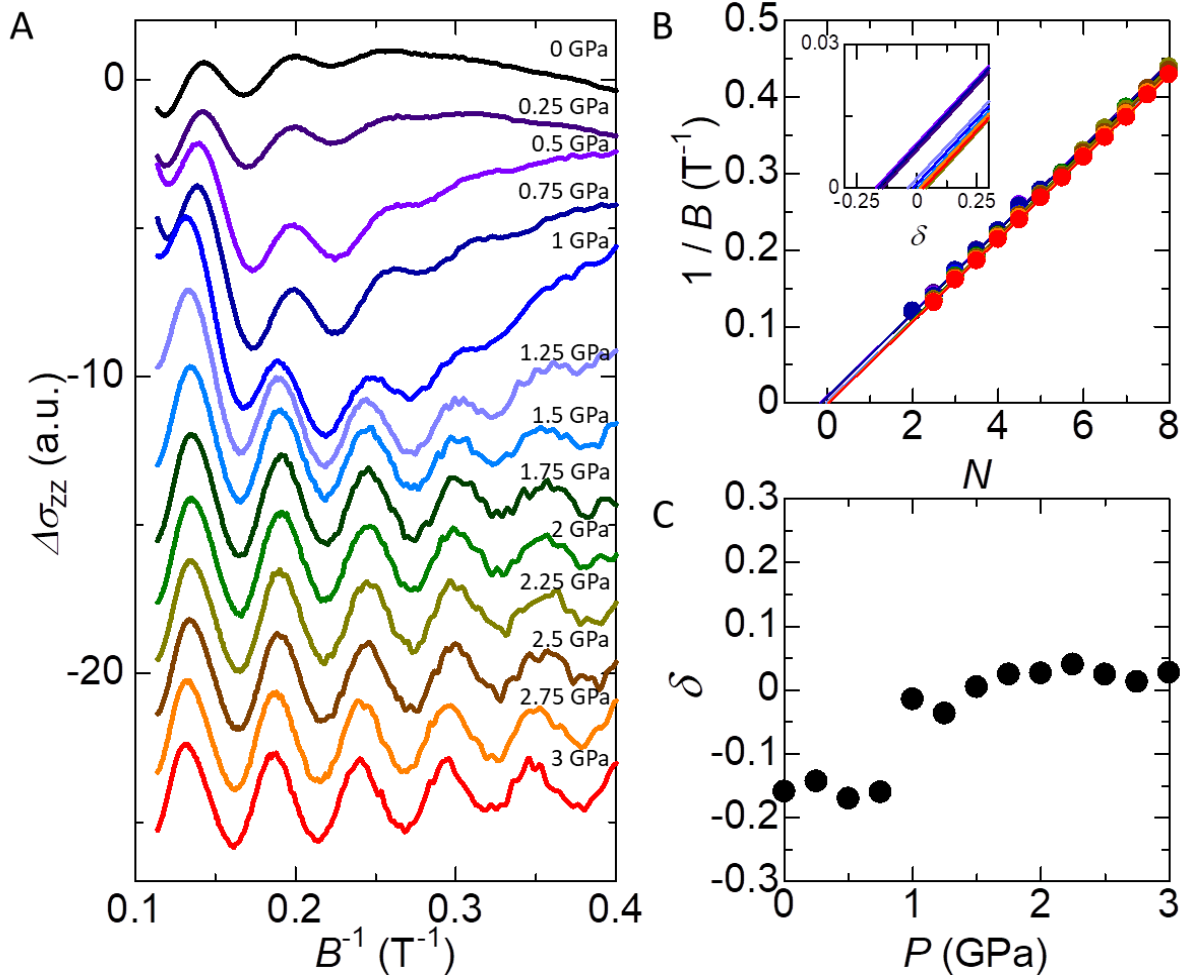

**Fig. S6.** $\sigma_{zz}$ oscillations of sample 2. (*A*) SdH oscillations of $\sigma_{zz}$ in sample 2. (*B*) Index plot for $\sigma_{zz}$ oscillations in sample 2. (*C*) Pressure variation of the horizontal axis intercept $\delta$ in the index plot. $\delta$ shows the similar crossover behavior toward *P* = 2 GPa as that for $\rho_{zz}$ oscillations.

## 6. SdH oscillations in sample 3 and 4

Anomaly of the Fermi surface cross-sectional area $S_F$ has been also observed in sample 3 and sample 4. In Fig. S7, we show the oscillating components of sample 3 (Fig. S7 A) and sample 4 (Fig. S7 B) under various pressures. $S_F$ calculated from the period of oscillation as a function of pressure are displayed in Fig. S7 C (sample 3) and Fig. S7 D (sample 4), respectively. According to the oscillation periods, the Fermi levels of both samples are estimated to locate below $E_c$ (See above section 1). Therefore, pressure-induced Lifshitz transition which has been observed in sample 1 cannot be expected for both samples similarly to sample 2. Importantly, however, $S_F$ of both samples still shows anomaly around $P = 2$ GPa, which almost coincides with the theoretically predicted pressure for the topological phase transition; it shows the decreasing behavior toward $P = 2$ GPa and increases afterwards. These results indicate that measurement of SdH oscillations is a powerful probe for topological phase transition in Te even if Fermi level is slightly away from the valence band maximum.

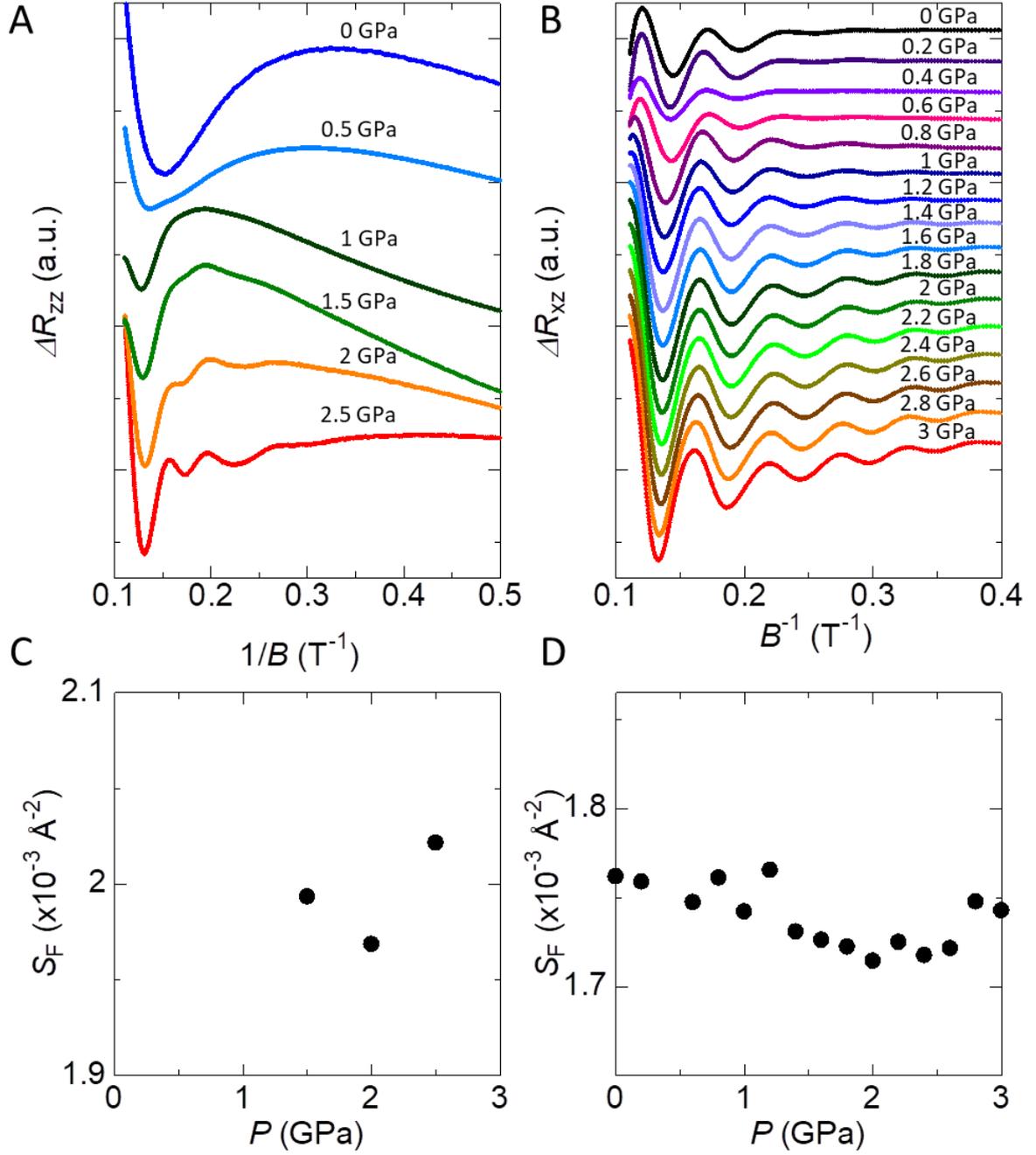

**Fig. S7.** SdH oscillations in sample 3 and 4. (*A* and *B*) SdH oscillations of sample 3 (A) and sample 4 (B). (*C* and *D*) Pressure dependence of the cross-sectional area of Fermi surface ($S_F$) estimated from the periods of oscillations. $S_F$ of both sample 3 (C) and sample 4 (D) show similar anomaly around $P = 2$ GPa as that of sample 2 in the main text.

## 7. SdH oscillations in sample 5 and 6 ($B \parallel I \parallel z$)

We further studied pressure dependence of SdH oscillations of Te in another configuration ($B \parallel I \parallel z$).

Figs S8 and S9 show the magnetoresistance (A), oscillating components of the resistance (B), index plots (C), and pressure dependence of the cross-sectional area of Fermi surface (D) for sample 5 (Fig. S8) and sample 6 (Fig. S9), respectively. In these samples, we applied both current and magnetic field parallel to the $z$ axis. Judging from the values of $S_F$, Fermi level of sample 5 and 6 are considered to locate below $E_c$. ($S_F$ =0.16× $10^{-3}$ $Å^{-2}$ when $E_F = E_c$ = -2 meV, which is smaller than those of sample 5 and 6.) In this configuration, periods of SdH oscillations reflects the maximal cross section of dumbbell-shaped Fermi surface which is illustrated by solid line in Fig. S10. This area is not sensitive to the topological phase transition which mainly affects the neck position drawn by dashed line in Fig. S10. Actually, anomaly of $S_F$ around $P$ = 2 GPa is small and hard to see (Figs. S8 D and S9 D). Instead, $S_F$ of both samples show the clear upturn around $P$ = 0.6 GPa, which coincides with the pressure at which camel-back-like valence band dispersion disappears. It is considered that disappearance of the camel-back-like valence band dispersion can be more sensitively probed in this configuration than the configuration discussed in the main text ($B \perp I \parallel z$).

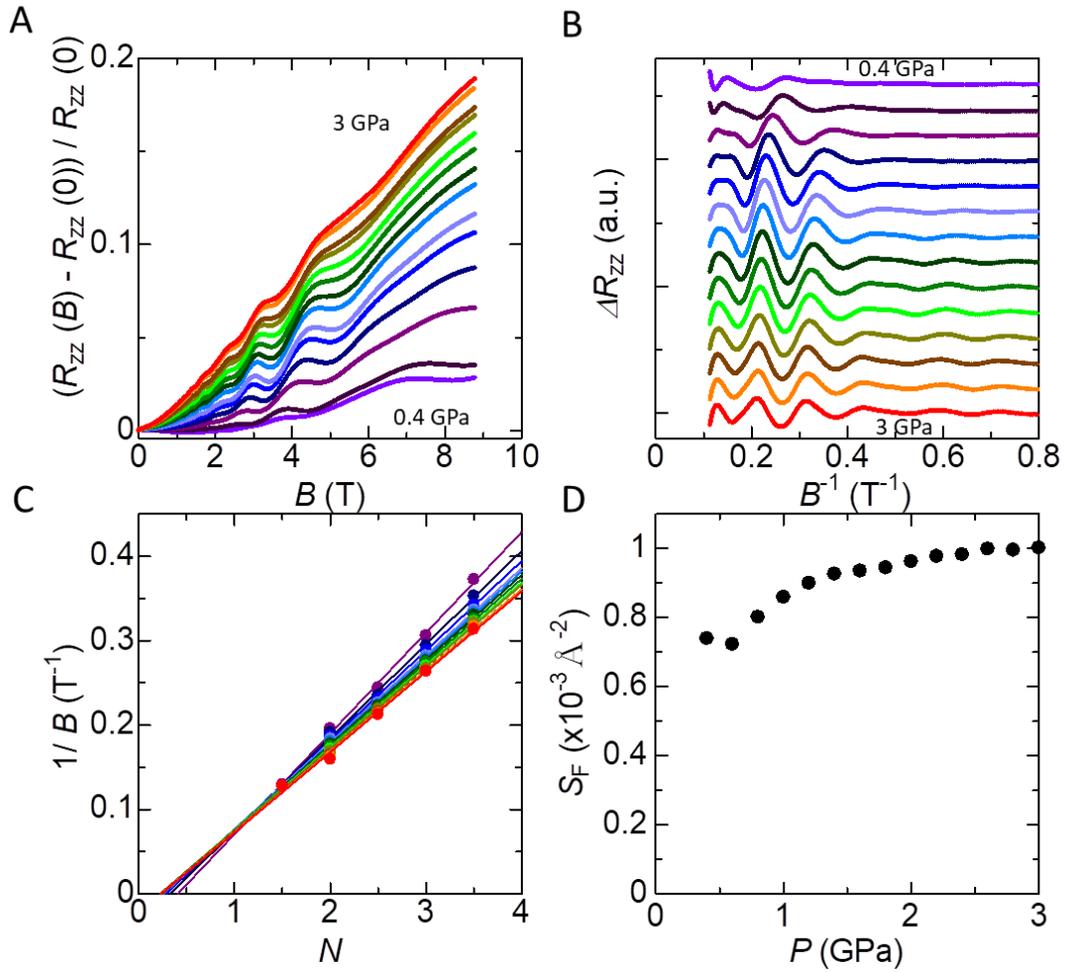

**Fig. S8.** SdH oscillations of sample 5 ($B \parallel I \parallel z$). (*A* and *B*) Magnetoresistance (A) and oscillating components (B) of the resistance calculated by subtracting the polynomial background from A. (*C*) Index plots for each pressure. (*D*) Pressure dependence of the cross-sectional area of Fermi surface estimated from the periods of oscillations.

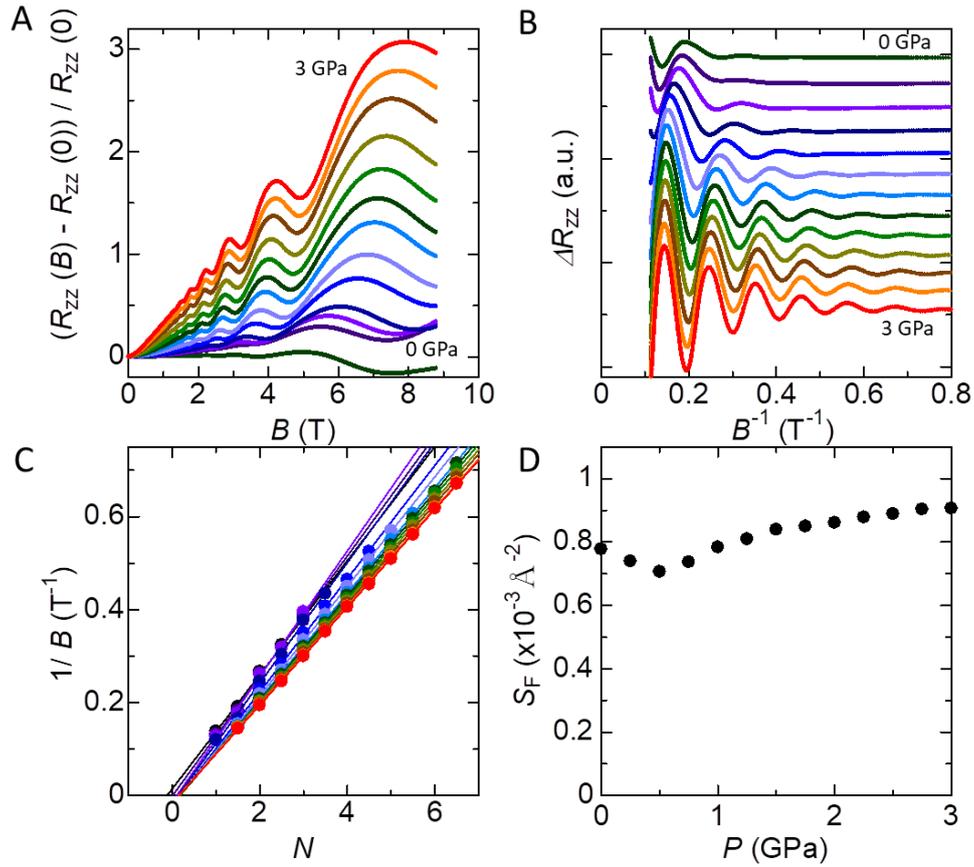

**Fig. S9.** SdH oscillations of sample 6 ($B \parallel I \parallel z$). (*A* and *B*) Magnetoresistance (A) and oscillating components (B) of the resistance calculated by subtracting the polynomial background from A. (*C*) Index plots for each pressure. (*D*) Pressure dependence of the cross-sectional area of Fermi surface estimated from the periods of oscillations.

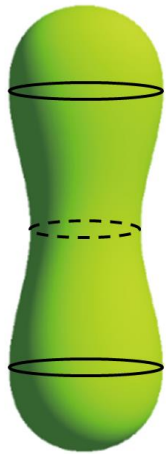

**Fig. S10.** Fermi surface and maximal (solid lines)/ minimal (dashed line) cross sections.

36